\documentstyle[preprint,aps,floats]{revtex}
\tightenlines
\begin{document}
\def\ba{\begin{eqnarray}}
\def\ea{\end{eqnarray}}
\def\be{\begin{equation}}
\def\ee{\end{equation}}
\def\({\left(}
\def\){\right)}
\def\[{\left[}
\def\]{\right]}
\def\lagrange {{\cal L}}
\def\del {\nabla}
\def\d {\partial}
\def\Tr{{\rm Tr}}
\def\half{{1\over 2}}
\def\fourth{{1\over 4}}
\def\bibi{\bibitem}
\def\S{{\cal S}}
\def\xx{\mbox{\boldmath $x$}}
\newcommand{\labeq}[1] {\label{eq:#1}}
\newcommand{\eqn}[1] {(\ref{eq:#1})}
\newcommand{\labfig}[1] {\label{fig:#1}}
\newcommand{\fig}[1] {\ref{fig:#1}}
\def\gsim{ \lower .75ex \hbox{$\sim$} \llap{\raise .27ex \hbox{$>$}} }
\def\lsim{ \lower .75ex \hbox{$\sim$} \llap{\raise .27ex \hbox{$<$}} }
\newcommand\bigdot[1] {\stackrel{\mbox{{\huge .}}}{#1}}
\newcommand\bigddot[1] {\stackrel{\mbox{{\huge ..}}}{#1}}
\title{Why Quantum Mechanics is Hard to Understand} 
\author{Doug Bilodeau\thanks{email:bilodeau@iucf.indiana.edu}}
\address{Indiana University Cyclotron Facility}
\address{2401 Milo B. Sampson Lane}
\address{Bloomington, IN 47408 USA}
\date\today 
\maketitle

\begin{abstract}

	To understand the foundations of quantum mechanics, we have to
think carefully about how theoretical concepts are rooted in --- and
limited by --- the nature of experience, as Bohr attempted to show.  
Geometrical pictures of physical phenomena are favored because of their
clarity.  Quantum phenomena, however, do not permit them.  Instead, the
historical and dynamical aspects of description diverge and must be
expressed in different but complementary languages.  Objective historical
facts are recorded in terms of objects, which necessarily have an
imprecise, empirical quality.  Dynamics is based on quantitative
abstraction from recurring patterns.  The ``quantum of action'' is the
discontinuity between these two ways of looking at the physical world.

\end{abstract}
\vskip .3in

\section{INTRODUCTION}

	There was a time when the Copenhagen Interpretation was commonly
thought to have solved the conceptual problems of quantum mechanics.  
That view is much less fashionable now, but we do not appear to be
approaching a consensus on any alternative.  Most papers written on this
subject start with the mathematical formalism (its predictions so
thoroughly validated by experiment) and try to recast it into some new
form with a natural and unambiguous physical meaning.  These efforts seem
to me to take us in the wrong direction --- deeper into abstraction (where
it is easy to mislead ourselves) and farther from the immediate empirical
basis of physical concepts, which is where the difficulty lies.

	I think Bohr was right, or nearly so.  Unfortunately, his writings
are so obscure that there is no more unanimity on ``what Bohr really
thought'' than there is on what quantum mechanics is all about.  I have
tried here to follow what I take to be the spirit of Bohr's analysis, but
in language which will be clearer to physicists today.  But not very clear
--- quantum mechanics is intrinsically hard, because it pushes us to the
limits of the very idea of the physical.

	Some linguists claim that the capacity for language is part of our
genetic makeup.  Even the grammatical structure of language is to some
extent biologically determined, e.g.,~the distinction between nouns and
verbs \cite{pinker}.  Perhaps our brains also favor certain kinds of
mathematical and physical concepts.  The machines we design and the
theorems we prove reflect the hard-wired structure of our thinking.
Whether this is true or not, it is true that quantum theory startles us by
distinguishing two aspects of the way we think about the physical world
--- two aspects we are used to describing in a single unified account:
what a thing is and what a thing does.  The result is that we cannot form a
single consistent picture of physical reality at the quantum level based
on geometrical structures in physical space and time.

	When Einstein wrote his first paper on special relativity, he used
mathematics already well known in contemporary research on
electrodynamics.  His profound and original contribution was to show how
physical concepts are based on the nature and limitations of observations,
and how intuitively appealing concepts like simultaneity are abstractions
which may or may not apply to the real world.  To understand quantum
mechanics, we also have to wrestle with physical concepts.  In this brief
essay, I cannot treat any particular examples in detail, but I have tried
to point out the crucial features of physical analysis which take on new
meanings as we go from classical to quantum physics.  Needless to say, my
discussion here is neither as original nor as revolutionary as Einstein's
was. I have for the most part restated very old ideas in slightly new
language, but hopefully new and fresh enough to encourage others to see
Copenhagen in a new light.

	In this paper I present the conceptual system which I think gives
the clearest understanding of what it is we are doing when we use quantum
mechanics.  I do not specifically argue against other views, such as
Bohmian mechanics or various versions of decoherence, etc.  My criticisms
of these ideas I reserve for possible future publications.  I assume that
the reader is familiar with elementary quantum mechanics, and also with
some of the classic literature on conceptual issues, e.g.,~as collected in
\cite{wz}.

\section{OBJECTS VS. THEORY}

	Physics (like all science) begins with physical objects, a
category which is indispensable and familiar but still somehow strangely
vague. Objects are things which are generally in constant interaction with
their environments.  We can continue to identify them through many sorts
of motions and changes.  We can observe them repeatedly and act on them
and observe the results.  Other people can observe all this along with us.
Factual information about objects is in principle verifiable by anyone.
Objects are persistently identifiable in this way (at least for a while),
but they also change, sometimes unpredictably, sometimes in ways hard to
detect or describe.  Heraclitus said we cannot step into the same river
twice --- i.e.,~it is never quite the same.  An object is like a river of
events and appearances which we see as a persisting unity.

	Objects are defined by indicating them --- e.g.,~``That big rock
over there next to the tree.'' I don't have to know everything there is to
know about the rock to specify it uniquely.  There may not even be a ``set
of all facts'' about an object.  Is mud caked on the side of the rock part
of it, or a separate object?  Objects are full of ambiguities like that
which we can never completely define away.  The important thing is that it
can be located within the network of familiar objects and events --- the
history of our world. This is in contrast to the abstract entities of
geometry.  All the properties of an equilateral triangle are implied by
its definition (given the axioms of the geometrical context in which it is
considered).  A physical object can always surprise us. The big rock by
the tree might turn out to be a hollow prop or a heavy iron meteorite.  
This contingent quality of genuinely physical things can be annoying to a
theorist, because it reminds us that all theoretical descriptions are
necessarily approximations and simplifications.  But we cannot describe
the world without at some point referring to objects --- it is only
through them that we have a handle on the physical world.  They pretty
much constitute what we mean by the physical world.  The continuity and
verifiability of objects help us distinguish ``objective'' reality from
fantasy.

	Physical thinking begins with physical objects, but physical
theory quickly builds up a very different class of concepts for dealing
with phenomena (e.g.,~space-time, point particles, trajectories, momentum,
energy, entropy).  The shift is obvious in a comparison of the thinking of
Aristotle and Galileo.  Aristotle still focuses on objects as such and
their essences, transformations, and causes.  This makes him very obscure
for a modern scientifically trained reader.  Galileo and his successors
look instead for descriptions in terms of quantifiable properties of
objects, and especially quantifiable relationships between them (such as
relative positions and velocities). Our beautiful and precisely tested
theories of matter refer to an abstract description of phenomena which is
made possible by restricting our attention to quantifiable properties.  
This realm of abstraction is one step removed from the concrete reality of
things.

\section{GEOMETRY VS. DYNAMICS}\label{sec:geodyn}

	Geometrical objects are static.  The mind's eye can scan over
them, making comparisons or measurements back and forth repeatedly as much
as we like. The fundamental geometrical ideas are congruence and measure.  
Physical processes and objects are dynamical (a word which implies both
change and a chain of causal links among objects). A physical event
happens once and can never recur --- only an event of the same {\it type}.
In physics we observe the consequences of events by means of our own
transient subjective experiences and by repeatable and objective reference
to objects whose properties have been affected.  Physical understanding
means knowing how to categorize types of events and the types of causal
connections which relate ``typical situations'' (based on observing or
setting up relevant conditions and ignoring irrelevant ones) to their
possible outcomes.  We as observers and experimenters are not external to
the physical world. All our actions are part of the web of events and
causal relationships. Physical concepts are based on the dynamical
properties of objects (including us).

	When we are accustomed to using a particular physical theory, the
process of connecting concrete objects with abstract theory seems very
straightforward.  The difference between those two things (concrete and
abstract) is really very subtle and full of pitfalls.  It is convenient to
talk as if our theoretical structures were ``isomorphic'' to the objects
out there in the world which they describe, but true isomorphism relates
one mathematical entity to another.  Only when a physical object can be
treated like a geometrical object --- e.g.,~a machined piece of metal
which we can inspect and measure repeatedly, examining it with the
physical eye in the same way the mind contemplates a geometrical figure
--- only then can we find something like a quantitative isomorphism.  
What we have then is a relation between geometrical measure and physical
acts of measurement. The latter are in a sense outside of time in this
case.  The results we get do not depend on when we do the measurements or
in what order.  Usually, physical analysis is not so simple.

	The static geometrical view of the physical world treats time as
essentially equivalent to space, as one coordinate of a space-time in
which the history of the universe exists as a single geometrical object.
The units of analysis are geometrical objects --- lines, areas, and
volumes which indicate the trajectories of objects or the space they
occupy.  This simple view works at the classical level.  Ultimately,
however, we are forced to form our concepts on the basis of the practical
units of physical analysis in a dynamical universe --- interactions
between objects.

\section{DYNAMICS VS. HISTORY}

Suppose we could have a ``complete'' history of the universe --- we know
exactly what has happened at every point in space-time.  But to say that
``q'' happened at point $({\bf x},t)$ is meaningless unless q is a member
of a set of the possible things which can happen, i.e.,~it is an abstract
property distilled from a knowledge of the range of patterns which occur
in the universe as a whole.  The way we sort out the {\it whens} and {\it
wheres} in physics is different from the way we sort out the {\it whats}.  
The distinction is almost imperceptible in classical physics.  A vector
from my own position to the location of a planet determines the
gravitational force which the planet's mass exerts on me, but the vector
also tells me where to look to find it in the sky.  In quantum theory,
however, it becomes clear that locations in space and time take their
meaning from the concrete world of objects, but the way we categorize
properties comes from dynamics.  These are two irreducible aspects of
physical description which evolve together --- historical and dynamical.

	A thing is historical insofar as it is objective (can be observed
and treated as an object).  It then enters into the realm of recordable
objective occurrences which can be ordered in historical space and time.  
It is dynamical insofar as it is defined as an abstract element of a
dynamical theory which explains causal relationships between objects. A
quantum entity can be both (a common source of confusion).  A proton is an
object if we are talking about its track through a system of detectors.
More often the proton concept appears as an abstract constituent of a
dynamical system.

	Imagine that we could see the universe as omniscient external
observers, all of space and time at once, and that what we ``see'' is a
tangle of intersecting particle world-lines (cf.~Ch.~1 of Misner, Thorne,
and Wheeler \cite{mtw}).  We might detect some patterns which would
constitute physical rules or laws in some sense, but it would be quite
difficult or impossible to know whether we had found all the important
patterns, or to distinguish significant relationships from accidental
ones.  Even more difficult would be to translate this omniscient
description into the kinds of relationships and laws which would be
observed by the huge clumsy bunches of world-lines which constitute
ourselves.

	When we set out to investigate Nature, we are not like that
external omniscient observer at all.  We look for relationships and
patterns in the behavior of objects we know.  We want to find out -- does
this kind of object always behave this way under these circumstances?  
The phrases ``this kind,'' ``this way,'' and ``these circumstances'' imply
the ability to abstract relevant or significant features from what are
really unique events.  They also imply that we can find or (even better)
set up many instances of these typical situations.  The result is that the
concepts we develop to describe physical phenomena depend not only on what
we can observe, but also on what we can do.

	To say that A affects or causes or influences or interacts with B
implies a counterfactual:  If A had been different, B would have been
different, too.  The most convincing way to establish a connection is to
``wiggle'' some parameter in A more or less randomly and then observe the
same odd pattern showing up in some property of B \cite{newton}. If I
want to know whether a wall switch controls a certain light, I can flip
the switch on and off and observe whether the light follows my actions.
There is always the possibility that the light is being controlled by
someone else or goes on and off spontaneously; but if I put the switch
through a very irregular and spontaneous sequence of changes and the light
still follows along, then the probability of a causal connection is very
high (barring a conspiracy to deceive the experimenter).

	Physical theory is possible because we {\it are} immersed and
included in the whole process --- because we can act on objects around us.  
Our ability to intervene in nature clarifies even the motion of the
planets around the sun --- masses so great and distances so vast that our
role as participants seems insignificant.  Newton was able to transform
Kepler's kinematical description of the solar system into a far more
powerful dynamical theory because he added concepts from Galileo's
experimental methods --- force, mass, momentum, and gravitation.  The
truly
external observer will only get as far as Kepler.  Dynamical concepts are
formulated on the basis of what we can set up, control, and measure.

\section{REDUCTION IS DYNAMICAL, NOT GEOMETRICAL.}

	Human beings are adept at contriving machines which we assemble
part by part.  We try to understand nature by the inverse process of
dissection into component parts.  Twentieth century physics has probed
phenomena at ever smaller scales and we find that matter can be carved
into more or less fundamental pieces --- atoms and particles.  These
pieces are nothing like the hard bits of matter Newton or Descartes might
have expected, nothing like the machined metal part I mentioned in
Section~\ref{sec:geodyn} above.  This should not be surprising, because
physical dissection of objects is not like the intellectual dissection of
geometrical entities. Machines are an exception because they are designed
--- they are intellectual patterns imposed on matter.

	We probe matter by contriving interactions and studying dynamical
patterns.  The end result is not building blocks of matter, but
fundamental units of interaction --- quanta.  An exchange of one quantum
is the minimal dynamical relationship.  It is the simplest way one object
can interact with another.

	Quantum particles are not mechanical components.  In phenomena
where classical approximations are appropriate, it is useful to think of
atoms and particles in that way.  At a finer scale, it soon becomes
apparent that our mechanical concepts are not fundamental.  In a sense
Plank was correct to say that the electromagnetic radiation in a cavity is
not composed of photons, but that they are merely the discrete units of
energy which the field exchanges with its environment.  The field itself
is not discrete --- composed of particles.  It is simply ``quantized.''  
Quantum fields are not constitutive but dynamical descriptions.  They tell
us not what the field is but what it can do.

\section{CONFIGURATION SPACE VS. PHYSICAL SPACE}

	Newtonian mechanics combines the historical and dynamical aspects
of physics into one unified description.  Any object is made of particles
with trajectories ${\bf x}_i (t)$ (where the $i$ indexes the set of
particles).  These can be thought of as sequences of historical events (a
certain particle being in a certain place at at certain time) in the
concrete physical world we live in.  The same numbers ${\bf x}_i$, along
with inherent particle properties such as mass and charge, also determine
all the forces at work in the system --- i.e.,~the dynamics.  This way of
thinking about the physical world is so neat and appealing, it is natural
to regard it as an ideal to which physics should conform as much as
possible.  It may even be difficult to imagine that a physical theory
could be anything else.  But the quantum method of describing phenomena is
radically different.

	In the familiar Schr\"odinger approach, we have not trajectories
but a wave function $\psi({\bf x}_i ,t)$.  Now the ${\bf x}_i$ are
independent variables along with the time.  This looks at first like a
field theory, but $\psi$ cannot be given a realistic historical
interpretation in the way that the electromagnetic field can.  The ${\bf
x}_i$ represent not historical space (the realm of objective events) but
the configuration space of the dynamical situation we are analyzing (the
possible degrees of freedom of the interaction).  The space of states is
the space of well-behaved complex functions on the configuration space.  
The wave function description presupposes a prior objective context.  We
partition the objective world into chunks with carefully prepared
properties (e.g.,~source, target, and detector).  The wave function
represents the dynamical relationship among these objects.  That is a
crucial point and the most common stumbling block.  The quantum ``system''
is usually something small --- one or a few particles.  The system gives
us the degrees of freedom of the dynamics.  But the dynamics is a causal
relationship among big things --- the parts of the apparatus.  The quantum
experimenter goes to a lot of trouble to make big things interact in small
ways, so that the discreteness of the quantum of action shows up in the
behavior of objects we can experience directly.

	From $\psi$ we calculate probabilities of various responses of the
detector in our experimental setup (scattering cross-sections, branching
ratios, etc.). The wave function or state vector encompasses many possible
outcomes, so if we try to take it out of its dynamical context and give it
a historical interpretation, then we may imagine as a consequence ``many
worlds'' or ``collapse'' events.  This comes of confusing the
configuration space with historical space.

	How {\it are} these two spaces related, if they are not identical?
I suggest that it is useful to think of the configuration space as
something like a tangent space (a crude metaphorical use of the
mathematical term). It is not tangent to a point of historical space-time,
but rather is associated with a particular ``dynamical partition'' of the
physical world, i.e.,~a certain way of dividing the world into objects.

	A quantum state is a little like the velocity vector of a flow on
a manifold.  We require different tangent spaces to describe the velocity
at different locations.  Similarly, when we alter the conditions of an
experiment or perform a measurement or otherwise gain objective
information, we shift to a new dynamical partition, so that there is a
discontinuity between the old state description and the new one
appropriate to the new context.  Perhaps it is no coincidence that the
Schr\"odinger equation is linear even though quantum transition phenomena
have a nonlinear flavor --- it may be (again speaking metaphorically)
because the Schr\"odinger equation is defined on a linear tangent space
rather than on the full historical manifold.

	I doubt that there is a well-defined set of all possible dynamical
partitions.  If there were, then perhaps we could arrive at a realistic,
abstract, historical description of quantum dynamics by integrating
(another metaphorical abuse of a mathematical term) the states in the
configuration spaces over the set of partitions.  But even if some such
procedure actually produced a meaningful result, it still might not be
useful or important.  The mathematical structure might be too remote from
anything we could observe or measure to be intelligible.

\section{COMMON MISCONCEPTIONS}
 
	I want to comment briefly on three errors sometimes found in
discussions of quantum mechanics, often as a misrepresentation of Bohr's
explanation of the theory by those attacking or even those defending him.

\begin{description}

	\item[Error 1:] ``Two theories are needed in physics: classical
	physics for macroscopic and quantum mechanics for microscopic
	phenomena.''

\end{description}

	Bohr's terminology was a little unfortunate when he talked about
the ``classical'' nature of the measuring apparatus.  I think what he
meant was that physics develops as a refinement of our ordinary experience
with objects.  The mathematical concepts of quantum theory are dynamical,
not ontological, and we still need the objective aspect of historical
description to give them meaning --- the same objective aspect which was
indistinguishable from the dynamical aspect of classical physics.  Now we
still use the same historical objective concepts to describe an experiment
or observation even though the dynamical language is entirely different.
For practical purposes, we use classical physics where it works, but there
is no reason to think that a fully quantum analysis of any dynamical
system would not give the correct answers if we could do the calculations.

	Sometimes we use the geometrical structure of physical description
to sharpen and criticize our ideas about what is really out there and what
really happens; e.g.,~a~pencil in a glass of water is not really broken at
the water's surface -- the appearance is due to the refraction of light by
water; similarly a rainbow is not really a multicolored ribbon spanning
the sky -- it is a trick of sunlight refracting through water droplets
suspended in the air.  This is the kind of refinement of description
physics has always pursued.  Newtonian mechanics made it appear reasonable
to many that all phenomena could ultimately be reduced to an underlying
geometrical structure of matter which would be the one true description of
what was ``really there.''  Quantum physics has given up looking for that
kind of geometrical ontology.  We refine our description of objects as
precisely as we can, and then quantum mechanics gives us the dynamical
relations.  There is no ground floor of quantum description which replaces
the pragmatic description of ordinary objects.  We have atoms and
elementary particles which are useful for describing dynamical systems,
their conserved quantities and degrees of freedom, but these are not
fundamental in a historical/ontological sense.  They are not mechanical
components with individual well-defined locations and causal roles at
every moment of time.  This becomes obvious when we look at field
theoretical descriptions of the quantum vacuum.

	That we still need to base our description of quantum phenomena on
observation of macroscopic objects is not a weakness or dilution of
quantum mechanical principles.  That was the whole idea of quantum
dynamics from the very beginning when Heisenberg decided to represent
dynamical variables by operators.  When Bohr insisted that the quantum
revolution was irreversible, that we could never return to classical
physics by means of hidden variables or any other trick, he was affirming
among other things that our experience with atomic physics had revealed
what a sufficiently perspicacious natural philosopher might have realized
anyway, that the historical and dynamical modes of description are
independent (but complementary) and may take different forms.  The quantum
of action is the natural fault line between these two ways of looking at
things.

\begin{description}

	\item[Error 2:] ``There are two kinds of quantum processes: the
	continuous evolution of quantum states according to the
	Schr\"odinger equation, punctuated by discontinuous collapse or
	projection events which occur when a measurement is made.''

\end{description}

	From classical physics, we get the habit of thinking of a state as
a state of being, independent of context.  A quantum state is a dynamical
relation defined within an objective context.  The continuous evolution of
the state is the evolution of interaction amplitudes based on prior
objective information.  A new measurement changes the context.  If we are
still considering after the measurement a dynamical relation mediated by
the same or same type of particle (or set of particles), then it might be
convenient to think of the new state as a projection of the prior state
onto a subspace determined by eigenvalues of the observable measured.  
Actually, as I suggested in the previous section, the two states are
defined on entirely different ``tangent spaces.''  Once this is
understood, the Measurement Problem and associated paradoxes disappear.
\footnote{When people talk about the measurement problem, they focus on
the last step in quantum dynamical analysis --- the translation from the
final quantum state (which may encompass many possible outcomes) to a
particular objective result of the measurement.  They forget about the
first step --- the translation from a unique objective situation to an
abstract initial quantum state based on properties we consider relevant
and can control or measure.  There is a discontinuity in mode of
description at both ends.  The initial state is no more an objective
ontological state of being than the final state.}

	Many of these paradoxes involve an infinite regression of
contexts.  Wigner's friend looks at his apparatus and then Wigner asks his
friend the result.  If ``measurement'' causes the state to collapse, is
Wigner's friend in a superposition of states until he gives his answer?
The superposition principle applies to dynamical states, not objects.  
The quantum state of the apparatus is generally irrelevant to the analysis
of an experiment.  We could in principle determine such a state, but first
we would have to specify an objective context in which the apparatus
functions as a carrier of dynamical influence.  There might be many ways
to specify the context, and each one result in a different state, but
without any difference in the objective physical reality.  It is not
really meaningful to talk about the state of an object unless it is
treated in our analysis as a dynamical system in an objective context.
Meanwhile, an object is what it is --- an object, not an abstraction.

	If the answer is that simple (and my version is not very different
from what many others have said or even from Bohr's explanation in 1927)
why is there still controversy after so many years? The emotional and
intellectual appeal of the static geometrical picture of space-time and a
mechanical view of matter should not be underestimated.  If the world is a
machine, then perhaps we can control it, especially if the watchmaker who
made it is blind and cannot interfere (to use the peculiar metaphor in the
title of a book by Richard Dawkins \cite{dawkins} --- can the world be a
mechanism without a design engineer, albeit an aimless one?)  If the
world is simply a collection of atoms in motion, then superstition is
without basis; there can be no transcendent evil to fear, as Lucretius
taught (but perhaps not much to value either). If the world is nothing but
mathematical form, then our minds can penetrate to the ultimate foundation
of things and we can escape the dreariness of mundane personal life by
contemplating ``the thoughts of God when he made the universe.'' Of
course, we can develop technology, rationally manage our environments,
reject superstition, and find joy in Platonic contemplation no matter what
physical theory happens to be true.  But the spell of geometry and
abstraction is powerful, and we can come to depend on it.

\begin{description}

	\item[Error 3:] ``Bohr's view of quantum mechanics implies a
	shadowy, mystical picture of the world in which the mind of the
	observer influences events in occult ways and there is no
	objective independent physical reality.''

\end{description}

	Classical physics encouraged the belief that whatever cannot be
described mathematically does not exist.  To Locke and others, the primary
attributes of objects were geometrical.  The physical world is simply a
geometrical form realized in matter.  Attributes or qualities which could
not be described in this way were secondary -- not existing in themselves
but only in the subjectivity of the beholder.  The more militant version
of this view is that physical entities must not only be describable
mathematically, they must {\it be} mathematical constructs. The universe
must be founded on a mathematical ontology.  To some, this became the
meaning of ``objective independent physical reality.''

	I think only students who have invested a lot of time and effort
studying classical mechanics are likely to be shocked to learn that
quantum theory does not have such an ontology.  It has only been since the
1970s that the general public (or even many physicists) had any idea there
was a Problem of Reality in quantum physics.  The indeterminacy of quantum
dynamics made a little stir earlier in the century because of a possible
connection with the classical philosophical problem of free will.
Otherwise, relativity got all the press as {\it the} weird theory of 20th
century physics.

In the 1950s it was received teaching that Bohr and von Neumann had solved
all the conceptual and mathematical problems of non-relativistic quantum
mechanics (cf. for example \cite[pp.~44--45]{bernstein}.  I doubt that
many physicists really understood Bohr's ideas or realized that Bohr's
treatment of measurement was radically different from von Neumann's.  
(Bohr would never have assigned a quantum state to the measuring
apparatus.)  Perhaps the shift in the center of gravity of physics
research from the Europe of Kant and Husserl to pragmatic North America
did not help.  Our physics departments are good at teaching formal
theoretical methods and experimental techniques, but how to understand the
connections between one set of concepts and the other is often left as an
exercise for the student.  Small wonder that the conceptual foundations of
quantum mechanics as given in textbooks in confused and perfunctory
accounts of the Copenhagen Interpretation looked to students like a closet
full of skeletons. The collision of rigid (because rarely examined)
orthodoxy with playful revisionism generated a cadre of popularizers, who
preached to the multitudes that quantum mechanics was strange in a very
cool way, and that in fact perhaps all strange and cool things were linked
to it.  A pop language was created which popular writers from very
different backgrounds were not slow to exploit.

	This is not necessarily a bad thing.  After Newton and the
industrial revolution, classical physical terminology often appeared in
common speech well out of its proper context, sometimes to legitimize
outright frauds or fallacies (animal magnetism), sometimes as useful
metaphor (social entropy, a team's momentum in sports).  Why not quantum
leaps of achievement and economic superfluidity?  Anything radically new
seems magical, and so it becomes a metaphor for the magical bursts of
insight which constitute the milestones and landmarks of human experience.

	It remains true nonetheless that objectivity, solidity, and
regularity are not compromised by quantum theory.  The objectivity of
physics comes from its grounding in experience.  No formalism can dissolve
or erase the basic truths of everyday experience.  Schr\"odinger's cat is
definitely alive or dead, even if the quantum formalism can describe only
its dynamical state, not its state of being.  Physics gives a complete
{\it dynamical} description of all phenomena --- i.e.,~all causal links
and transformations as measured physically.  That doesn't mean physics
explains all higher level emergent phenomena (consciousness, culture,
morality) or that it should be expected to.  Nor does it provide an
underlying ontology for reality in all its aspects \cite{bilodeau}.

	There are still some hard cases for which our experience with
microscopic processes gives us little help.  Quantum dynamics presumes an
objective context, but if we are interested in the dynamics of the
universe as a whole in an early epoch in which the quantum properties of
space-time itself dominate, then there is no conventional recipe for
understanding what is going on.  Perhaps the present is the objective
context for the past.  In any case, we should learn from past experience
to expect answers only to questions we can put to the universe with action
and observation, not with reason and imagination alone.

\end{document}